\documentclass[letterpaper,10pt]{article}   

\usepackage{LeGquant-ph,amssymb, amsmath, mathrsfs}
\newcommand{\ket}[1]{| #1 \rangle}
\newcommand{\Set}[2]{{\{ #1 ,\ldots, #2 \}}}
\newcommand{\Natural}{\mathbb{N}}
\newcommand{\abs}[1]{\vert #1 \vert}
\newcommand{\ceil}[1]{\lceil #1 \rceil}
\newtheorem{theorem}{Theorem}[section]
\newtheorem{corollary}[theorem]{Corollary}
\newtheorem{proposition}[theorem]{Proposition}
\newtheorem{definition}[theorem]{Definition}
\newtheorem{fact}[theorem]{Fact}  
\newenvironment{proof}{%
  \noindent{{\bf Proof}.}}{%
  \hspace*{\fill}$\square$
  \vspace{2ex}}
\newenvironment{proof2}{%
  \noindent{{\bf Proof of Theorem 3.4}.}}{%
  \hspace*{\fill}$\square$
  \vspace{2ex}}
\begin{document}   

\vspace{14mm}
\begin{center} 
{\LARGE \bf Exponential Separation of Quantum }\vspace{2mm}

{\LARGE \bf and Classical Online Space Complexity}\vspace{8mm}  

{\large Fran{\c c}ois Le Gall}\vspace{6mm}

{\it Department of Computer Science, The University of Tokyo}\\
{\it 7-3-1 Hongo, Bunkyo-ku, Tokyo 113-0033, Japan} \vspace{1mm}

and\vspace{1mm}   

{\it ERATO-SORST Quantum Computation and Information Project}\\
{\it Japan Science and Technology Agency}\\
{\it 5-28-3 Hongo, Bunkyo-ku, Tokyo 113-0033, Japan}\vspace{3mm}

email: legall@qci.jst.go.jp\vspace{8mm}

\setlength{\baselineskip}{9.5pt}
      \begin{quotation}
\noindent{\bf Abstract.}\hbox to 0.5\parindent{}

Although quantum algorithms realizing an exponential time speed-up over the best known classical algorithms exist,
no quantum algorithm is known performing computation using less space resources than classical algorithms. 
In this paper, we study, for the first time explicitly, space-bounded quantum algorithms
for computational problems where the input is given not as a whole, but bit by bit.
We show that there exist such problems that a quantum computer can solve using exponentially less work space than a classical computer.
More precisely, we introduce a very natural and simple model of a space-bounded quantum online machine
and prove an exponential separation of classical and quantum online space complexity,
in the bounded-error setting and for a total language.
The language we consider is inspired by a communication problem (the disjointness function) that Buhrman, Cleve and Wigderson used 
to show an almost quadratic separation of quantum and classical bounded-error communication complexity.
We prove that, in the framework of online space complexity, the separation becomes exponential.
\end{quotation}
\end{center}


\section{Introduction}   


Space complexity studies the amount of work space necessary to solve computational problems. 
In particular, one of the most active research topics is sublinear space complexity, the study of computation 
where the amount of work space available is less than the size of the input.
In this case, work space has to be separated from the input space. This is usually done by supposing
that the input is written on a read-only memory and that
an additional read-write work memory is available.
Sublinear space complexity is thus practically meaningful in the case of computational 
devices for which the cost of work memory is prohibitive compared with the cost of (read-only) input memory.
One of the main reasons why
the study of quantum sublinear space complexity is of paramount importance is that quantum computers
are precisely such kinds of devices.  
Indeed, defining quantum  space complexity is not an easy matter but, intuitively, the input
of the computation being classical, it is natural to consider that, even for a quantum computer, 
the input is stored on a classical memory. In this case, the quantum space complexity of a computational task 
corresponds directly to the  amount of quantum memory necessary for the computation.
Since, today, one of the main technological obstacles to the construction of a quantum computer seems to be 
the realization of quantum memory, studying what can be done with a small scale quantum memory is extremely important and
this is precisely what we study in this paper. 
Another connected  motivation for the study of quantum space complexity is the theoretical and practical issue of understanding
whether it is possible to design quantum algorithms that use less quantum work space 
than the classical work space needed by the best classical algorithms. 

Space complexity of quantum Turing machines has been studied by Watrous \cite{WatrousJCSS99, Watrous03}.
He first defined a model of quantum Turing machines allowed to perform 
a very restricted kind of measurements during the computation \cite{WatrousJCSS99}  
and proved the following rather surprising negative result:
in the unbounded-error setting 
(the computation has to be correct with probability strictly greater than 1/2, but this probability can
be arbitrary close to 1/2),
this model of quantum computation
is equivalent, as far as
space complexity is considered, to classical probabilistic computation.
Watrous then defined a far more general model of space-bounded quantum device \cite{Watrous03} and showed that
even this strong model is also equivalent, as far as unbounded-error
space complexity is considered, to classical probabilistic computation.
Using the result by Borodin, Cook and Pippenger \cite{Borodin+83} that states that any unbounded-error probabilistic computation
can be simulated by a deterministic machine with at most a quadratic increase of space, this implies
that any exact, bounded-error or unbounded-error 
quantum computation that uses $s$ quantum space resource can be simulated by a deterministic
classical computation using $s^2$ space. Thus the gap between space-bounded quantum and classical computation 
can at most be quadratic.


In this paper we show that,
when the input is given online, i.e.,bit by bit, and not as a whole, the situation
changes dramatically\footnote{ 
In this paper, the term ``online'' refers to the concept of online Turing machines.
We stress that this view of online computation is slightly different from the one usually adopted
when studying competitive analysis of online algorithms. }: there
are computational problems for which a quantum computer can use exponentially less space resources 
than a classical computer. 
This model of computation corresponds to the notion of streaming algorithms 
and is the model of choice for extremely long inputs that cannot be stored in memory,
like data from large databases or from the Internet
(we refer to \cite{Mut05} for a good survey of the active research field of classical streaming algorithms).

To obtain our separation,
we introduce a model of online quantum machine,
very simple and weaker than Watrous's model \cite{Watrous03}.
Informally, our quantum machine can be considered as a classical probabilistic Turing machine with an additional 
quantum register.
The particularity of this model is that the classical and quantum parts are completely separated, giving
a very realistic model. Indeed, if a quantum computer can be built, it will be
meaningless to ask it to perform tasks that can done efficiently using classical computation.

Our main result is an exponential separation of 
quantum bounded-error online space complexity 
and classical  bounded-error online space complexity
for a total language. 
The language we consider is based on the following online computational problem, inspired by a well-known
problem from communication complexity used by Buhrman, Cleve and Wigderson \cite{Buhrman+STOC98}. 
Two binary strings of the same length, $x=x_1\cdots x_m$ and $y=y_1\cdots y_m$, are 
repeatedly input, bit by bit: 
the computational device receives $x_1$, then $x_2$, up to $x_m$, then
$y_1$ up to $y_m$ and again $x_1$ $\dots$, the alternation being repeated about $\sqrt{m}$ times.
The problem is to decide whether there is an index $i\in\Set{1}{m}$ such that $x_i=y_i=1$.
If the classical device can store the two strings in memory, the problem is trivial but, in the case where
the length $m$ of the string is extremely large and is far beyond the capacity of the memory,
this approach is impossible.
We show, using arguments from communication complexity, that, for a classical bounded-error online Turing machine, it is indeed impossible to 
solve this problem with noticeable probability if less
than $\Omega(\sqrt{m})$ memory is available. 
Then we show that, for a quantum  bounded-error online machine, $O(\log m)$ memory (classical bits and quantum qubits) 
is sufficient to solve this problem with high probability, by simulating the protocol of \cite{Buhrman+STOC98}.
We use these results to give the first exponential separation of quantum and classical online space complexity,
for a total language and in the bounded-error setting. 

To our knowledge, quantum online space complexity has never been explicitly studied before this work\footnote{
But notice that, in the different context of finite automata, and not Turing machines, Ambainis and Freivalds 
have shown that quantum automata can recognize some languages using exponentially less states than any 
classical automaton \cite{Ambainis+FOCS98}.}. 
However, 
online space complexity is strongly connected with communication complexity.
In particular, any separation of quantum and classical one-way (only one-message sent) 
two-party communication complexity for a total function gives immediately,
under the assumption that the computational part of the quantum communication protocol can
be done space-efficiently, a separation of quantum and classical online space complexity classes.
From this observation, it is straightforward to show an exponential separation of quantum nondeterministic and classical nondeterministic
online space complexity from a result by de Wolf \cite{deWolfSIAMJC03}.        
But nondeterminism is an unrealistic model whereas our main result holds for the most realistic model of bounded-error computation.
Notice that some other separations are known for total functions in models different from the two-party setting, 
e.g.~the so-called simultaneous message passing model \cite{Buhrman+PRL01}, 
but there is no direct way to convert them into a separation in the online space complexity
setting. 

Separations of two-way (no restriction on the number of messages exchanged) 
quantum and classical communication complexity for total functions do not generally lead to a separation of
quantum and classical online space complexity.
However, the technique we use in this paper enables to obtain such a separation when
the quantum protocol has a particular form and 
the computational mode is sufficiently powerful to solve the string equality problem 
with small communication cost. These properties are 
satisfied by the quantum protocol involved in the separation of
quantum and classical bounded-error communication complexity by Buhrman, Cleve and Wigderson \cite{Buhrman+STOC98}, leading to our results.
Actually this technique can also be used to show a quadratic separation of classical nondeterministic
and quantum weakly nondeterministic online space complexity for a total language, using a result by Le Gall \cite{LeGallMFCS06}. 

We stress the following point: although the separation of \cite{Buhrman+STOC98} is only (almost) quadratic,
our separation in the framework of bounded-error online space complexity is exponential.
Our result, even though relatively simple, is indeed rather surprising: 
the best separation known of quantum and classical bounded-error communication complexity for a total function is quadratic 
\cite{Aaronson+05} and, in the framework of query complexity, it has been shown that classical and 
quantum bounded-error query complexities of total functions are polynomially related \cite{Beals+JACM01}.
We mention that an exponential separation for a partial function (a function with a promise on the inputs) in
the setting of bounded-error one-way communication complexity has been recently presented by Gavinsky, Kempe, Kerenidis, 
Raz and de Wolf \cite{Gavinsky+STOC07}. They used this separation to obtain an exponential separation of
quantum and classical online space complexity as well, but for a partial function. 
In comparison, our separation holds for a total function. 

Our result shows that there exist computational online problems for which the space saving can be exponential
over classical computers. 
From a practical point of view, this also shows that constructing a quantum computer with a
small scale quantum memory is sufficient to solve online problems that a classical computer with a large scale memory cannot solve.

Although the online computational problem we are considering in this paper is rather artificial, 
we hope that our work will be a first step in the direction of 
designing space-efficient quantum algorithms solving concrete problems for data streams.

The organization of this paper is as follows.
In Section \ref{section:spaceonline}, we recall definitions of online space complexity and present our model
of quantum online space complexity.
In Section \ref{section:spacebounded} we present our result:
an exponential separation of quantum and classical bounded-error online space complexity.

\section{Online Space Complexity: Definitions}\label{section:spaceonline}

\subsection{Classical online Turing machines}

We refer to, for example, Balc\'azar, Diaz and Gabarr\'o \cite{Balcazar+95} for a presentation of the general model of (offline) Turing machines.
Here, we explain the details of the model of online probabilistic Turing machines we consider in this paper and the notations we will use.

An online, or one-way, probabilistic Turing machine, that we abbreviate as OPTM, is simply a probabilistic Turing machine
with two tapes (the input tape and the work tape)
whose input tape is one-way, i.e., the tape-head can move to the right but never moves back to the left.
This means that the machine cannot go back to read the beginning of the input string and
has to store in the work tape all the information about the input string it will need in the future.
In the following, we will consider that the alphabet of both the input tape and the work tape is the ternary
alphabet $\Sigma=\{0,1,\#\}$.

The probability of acceptance of an OPTM $M$ on an input $w$, denoted $p_M(w)$,
is defined as the probability, over all coin flips of the machine, that the
input halts on an accepting state.
There are two ways of rejecting: either by stopping in a non-accepting state or by never halting.
Notice that, for bounded-error space complexity,
the condition that the machine halts on each input and each coin flips a priori changes the computational power of the machine.
The objective of this paper being to prove lower bounds on the space needed by classical machines,
we consider the strongest model, where the machine is not required to halt on each input. 

The space used for the computation on an input $w$ is the number of cells of the work 
tape used on the worst coin flips. 
We now define, for a function $s: \Natural \to \Natural$, 
the class of space-bounded
bounded-error online computation, denoted $OBPSPACE(s)$. 

\begin{definition}\label{def:OBR}
Let $L$ be a language over the alphabet $\Sigma=\{0,1,\#\}$. 
We say that $L\in OBPSPACE(s)$ if there exists an 
OPTM $M$ such that, for each $w\in\Sigma^\ast$:
\begin{enumerate}
\item[(i)]
if $w\in L$ then $p_M(w)\ge 2/3$;
\item[(ii)]
if $w\notin L$ then $p_M(w)<1/3$;
\item[(iii)]
$M$ uses at most $s(\abs{w})$ space on the worst coin flips.
\end{enumerate}
\end{definition}
The constants $2/3$ and $1/3$ are somewhat arbitrary, because amplification is possible (by running many computations in parallel)
up to any constant without increasing the space complexity by more than a constant multiplicative factor.
As usual, we will use notations of the form  $OBPSPACE(s(n))$ instead of $OBPSPACE(s)$. In this case, $n$ always represents the input length.

The configuration of an OPTM at a given step is the set of four elements
consisting of the current control state, the current positions of the two tape-heads and 
the current content of the work tape. 
The following fact is an immediate consequence of this definition.
\begin{fact}\label{fact3}
Let $M$ be an OPTM with tape alphabet $\Sigma$ and set of control states $Q$
such that, for each input $w\in\Sigma^\ast$, $M$ uses
at most $s(\abs{w})$ space on the worst coin flips.
Then the total number of different configurations that can appear
with positive probability during the computation by $M$ 
on inputs of length $n$ is at most
$$ns(n)\abs{\Sigma}^{s(n)}\abs{Q}.$$ 
\end{fact}

\subsection{Quantum online space complexity}
We refer to the textbook by Nielsen and Chuang \cite{Nielsen+00} for a good reference about quantum computation.

Designing a general model of quantum computation for the study of quantum space complexity
is not an easy task. 
Aharonov, Kitaev and Nisan \cite{Aharonov+STOC98} have shown that
the usual model of quantum Turing machines that allow measurements only at the end of the computation is
equivalent, with respect to time-resources, to
the most general view of quantum computation that includes measurements during the computation and thus is not unitary.
However, their simulation techniques cannot be used to prove equivalence with respect to space-resources.
Indeed, a direct simulation of a probabilistic computation by a quantum unitary machine need to remember
all the results of the coin flips and it thus terribly inefficient (a probabilistic computation using $s$ work space 
can use a number of coin flips exponential in $s$ and in the unitary quantum
computation model there is no way of reusing the space!). 
To obtain a definition of 
quantum space complexity such that quantum classes include classical probabilistic classes, 
it thus seems to be necessary to allow measurements during the computation.

The input being a classical string,
it is natural to consider that the input tape is classical. 
Watrous \cite{Watrous03} has defined such a model of ``hybrid'' quantum 
Turing machine, where measurements are allowed during the computation and their outcomes control the computation.
Actually, Watrous's definition is
very general and the interaction between 
the classical part and the quantum part of this machine is rather complex.  
In this paper, 
we consider a simpler, and weaker, version of quantum hybrid  machines, but still powerful enough to show 
separations of classical and quantum space complexity in the context of online computation (as we want to show
separations of classical and quantum computation, the weaker the quantum model is, the better). 
Informally, our machine can be considered as a classical probabilistic Turing machine that has access to a
quantum register, the initial state of this register being $\ket{0}\cdots\ket{0}$. Both the work space of the classical machine 
and the number of qubits of the quantum register that have been used during the computation are considered when defining the global space complexity.
This informal description is actually sufficient to understand the results of this paper: our upper bound (Theorem \ref{theorem:quantum})
is proved by presenting a space-efficient online quantum algorithm, and details about the model of the quantum online machine do not really matter.
Nevertheless, for completeness, we now describe carefully the model.

Denote by  $\mathscr{G}$ the set of gates $\{G_0,G_1,G_2\}$, universal for approximate quantum computation, 
where $G_0=H$ is the Hadamard gate, $G_1=T$ the $\pi/8$ gate and $G_2=CNOT$ the 2-qubits CNOT gate.
Let $\Sigma$ be a finite alphabet and $s$ be a function constructible using logarithm space such that $\Omega(\log n)\le s(n)\le n$. 
We are interested in sublinear complexity classes and, thus, we suppose that the input $w\in\Sigma^\ast$ 
(a classical string) is given as input to a classical OPTM.
Here the OPTM is seen computing a function: there is a third one-way write-only tape,  
on which the machine writes its output. Only the read-write work tape is taken in consideration in the definition of the space complexity of the OPTM.
We further suppose that $s(\abs{w})$ qubits are available, initialized to the state $\ket{0}^{\otimes s(\abs{w})}$.
These qubits are supposed to be ordered, each qubit labeled by an integer in $\Set{0}{s(\abs{w})-1}$.
Given two distinct integers $a$ and $b$ in $\Set{0}{s(\abs{w})-1}$, 
and any integer $i$ in $\Set{0}{2}$, we denote by $G_i^{[a,b]}$ the gate 
$G_i\in \mathscr{G}$ applied to the qubits number $a$ and $b$ 
(or only to the qubit number $a$ if $G_i$ is a one-qubit gate).
We use the convention that, if $a=b$, then $G_i^{[a,b]}$ represents the identity gate.
We now define formally the quantum online one-sided-error and bounded-error space complexity classes
that we denote respectively $OQRSPACE(s)$ and $OQBPSPACE(s)$. 
\begin{definition}\label{defbounded}
Let $L$ be a language over the alphabet $\Sigma=\{0,1,\#\}$. 
We say that $L\in OQRSPACE(s)$ [resp.~$L\in OQBPSPACE(s)$] if there
exists an OPTM $M$ such that, for each $w\in\Sigma^\ast$, the following holds:
\begin{enumerate}
\item[1.]
Whatever the outcomes of the coin flips are, $M$ halts after
at most $2^{s(\abs{w})}$ computational steps and uses at most $s(\abs{w})$ space.
\item[2.]
Whatever the outcomes of the coin flips are,
the contents of the output tape when $M$ halts is of the form
\begin{equation}\label{output}
a_1\#b_1\#c_1\#\cdots\#a_r\#b_r\#c_r, 
\end{equation}
for some integers $r\ge1$ and $a_i,b_i\in\Set{0}{s(\abs{w})-1}$, $c_i\in\Set{0}{2}$ $(1\le i\le r)$.
\item[3.]
If $w\in L$, then the output (\ref{output}) 
satisfies, with probability at least $1/4$ [resp.~at least $\sqrt{2/3}$], the following condition: 
the outcome of the measurement of the first qubit of the state $G_{c_r}^{[a_r,b_r]}\cdots G_{c_1}^{[a_1,b_1]}(\ket{0}^{\otimes s(\abs{w})})$
is 1 with probability at least $1/4$ [resp.~at least $\sqrt{2/3}$].
\item[4.]
If $w\notin L$, then the 
output (\ref{output}) 
satisfies, with probability 1 [resp.~at least $\sqrt{2/3}$], the following condition: 
the outcome of the measurement of the first qubit of $G_{c_r}^{[a_r,b_r]}\cdots G_{c_1}^{[a_1,b_1]}(\ket{0}^{\otimes s(\abs{w})})$
is 0 with probability $1$ [resp.~at least $\sqrt{2/3}$].
\end{enumerate}
\end{definition}
This definition corresponds to the following realistic model of quantum hybrid device, where
the computation consists in two stages.
First, the machine $M$ is used on the input string, giving a description of a quantum circuit $G_{c_r}^{[a_r,b_r]}\cdots G_{c_1}^{[a_1,b_1]}$.
Then, this circuit is applied to the quantum state $\ket{0}^{\otimes s(\abs{w})}$
and the first qubit of the resulting state is measured in the computational basis $\{\ket{0},\ket{1}\}$.
The input string is accepted if the outcome is 1, and rejected if the outcome is 0.
For a language $L$ in the class $OQRSPACE(s)$, if 
$w\in L$, then the probability that the string is accepted is at least $1/16$
(this value is somewhat arbitrary and can be increased by performing
amplitude amplification on both the classical and the quantum parts of the online machine); if 
$w\notin L$, then the probability that the string is rejected is 1. 
For a language $L$ in the class $OQBPSPACE(s)$, if 
$w\in L$, then the probability that the string is accepted is at least $2/3$; if 
$w\notin L$, then the probability that the string is rejected is $2/3$. 
Notice that in our definition,
we allow the same classical space and quantum space $s(\abs{w})$ for the computation. 

Practically, there is no need to store all the output of the $OPTM$:
the gates can be applied as soon as they are output. From condition 1., the total number of
gates applied is no more than $2^{s(\abs{w})}$. 
Moreover, with our definition,
the machine always halts and measurement of the quantum register 
is allowed only once, at the end of the computation. 
This is a weak
model compared with the general definition of quantum machines proposed by Watrous \cite{Watrous03}. However, we will show that such 
a model is still able to present an exponential save of space over the classical model.

Finally, we denote $OQRL=\bigcup_{c>0}OQRSPACE\:(c\log n)$ and 
$OQBPL=\bigcup_{c>0}OQBPSPACE\:(c\log n)$.

\section{Our Exponential Separation}\label{section:spacebounded}
In this section, we present
the exponential separation of classical bounded-error and
quantum bounded-error online space complexity.
Actually, we prove a stronger result: an exponential separation of classical 
bounded-error and quantum one-sided-error online space complexity.
\subsection{The disjointness problem}\label{section:searching}
We first shortly recall basic definitions of classical communication complexity.
We refer, for example, to Kushilevitz and Nisan \cite{Kush+97}
for further details.
Given a set of pairs of strings $X\times Y$, where $X\subseteq\{0,1\}^\ast$ and
$Y\subseteq\{0,1\}^\ast$, and a function $f:X\times Y \to \{0,1\}$,
the communication problem associated to $f$ is the following: 
Alice has an input $x\in X$, Bob an input $y\in Y$ and their goal is to compute
the value $f(x,y)$. We suppose that Alice and Bob have unlimited computation power.
In a randomized communication protocol, each player can flip (private) coins
and send messages according to the coin flip outcomes. 
We say that a randomized protocol is a bounded-error protocol for $f$ if, 
for each $(x,y)\in X\times Y$ the protocol outputs $f(x,y)$ with probability at least $2/3$.
The communication complexity of a bounded-error protocol $P$ 
that computes correctly $f$, denoted $R_{2/3}(P,f)$, is the maximum, over all
the inputs $(x,y)$, of the number of bits exchanged between Alice and Bob on this input on the worst coin flips.
The bounded-error communication complexity of the function $f$, denoted $R_{2/3}(f)$, 
is the minimum, over all the bounded-error protocols $P$ that compute $f$, of $R_{2/3}(P,f)$.
Quantum communication complexity is defined similarly, the only modification being that Alice and Bob can send quantum messages.
We refer to \cite{Buhrman00,Klauck00,deWolfTCS02} for good surveys of quantum communication complexity.

Now consider the following communication complexity problem, known
as Disjointness.\vspace{3mm}

$\phantom{aa}${\bf Disjointness} $\mathbf{(DISJ_n,\:\:n\ge 1)}$\vspace{2mm}

\indent $\phantom{aa}$Alice's input: a string $\mathbf{x}=x_0\cdots x_{n-1}$ in $\{0,1\}^n$\vspace{1mm}

\indent $\phantom{aa}$Bob's input:$\:\:$ a string $\mathbf{y}=y_0\cdots y_{n-1}$ in $\{0,1\}^n$\vspace{1mm}

\indent $\phantom{aa}$output: $\phantom{aaaaa}$$DISJ_n(\mathbf{x},\mathbf{y})=\bigwedge_{i\in\Set{0}{n-1}}((\neg x_i)\vee (\neg y_i))$\vspace{3mm}

\noindent 
In other words, $DISJ_n(\mathbf{x},\mathbf{y})=1$ if and only if there is no index $i\in\{0,\ldots,n-1\}$
such that $x_i=y_i=1$.
Buhrman, Cleve and Wigderson \cite{Buhrman+STOC98} have studied the quantum communication complexity
of $DISJ_n$ and shown the following result.
\begin{theorem}\label{theorem:upperbound}{\bf (\cite{Buhrman+STOC98})}
The quantum bounded-error communication complexity of the function $DISJ_n$ is $O(\sqrt{n}\log n)$.
\end{theorem}
Moreover, using the following well-known result of classical communication complexity, they notice that
this leads to an almost quadratic separation of quantum and classical bounded-error communication complexity.
\begin{theorem}\label{theorem:lowerbound}{\bf (\cite{Kalya+SIAMDM92,RazborovTCS92})}
$R_{2/3}(DISJ_n)=\Omega(n)$.
\end{theorem}
Notice that the result of Theorem \ref{theorem:upperbound} has been improved by H{\o}yer and de Wolf \cite{Hoyer+STACS02} 
and further by Aaronson and Ambainis \cite{Aaronson+05},
who gave a protocol using $O(\sqrt{n})$ quantum communication, which is optimal from a result by Razborov \cite{Razborov03}. 
This leads to a perfectly quadratic separation, but we will not use these results in this paper.

The quantum protocol of \cite{Buhrman+STOC98} realizing the upper bound is based on Grover's algorithm \cite{GroverSTOC96},
or more precisely its generalization proposed by Boyer, Brassard, H{\o}yer and Tapp \cite{Boyer+98}, 
that deals with the case where the number of solutions is unknown. It uses, on the
worst input and on the worst coin flips, at most $\ceil{\sqrt{n}}$ communication rounds, each message consisting
of $O(\log n)$ qubits. An important property of this protocol is that Alice and Bob have only to keep in memory the last message 
received  in order to compute the next message. 
This is the key observation leading to our result.
 
The total language we are considering is the following. 
\begin{definition}
Let $L_{DISJ}$ be the following language over the alphabet $\{0,1,\#\}$.
\begin{displaymath}
\hspace{-0mm}L_{DISJ}\!=\!\Bigl\{\:\:
1^k\#(\mathbf{x}\#\mathbf{y}\#\mathbf{x}\#)^{2^k}\!\big\vert\:\:\:k\ge 1; \:\:\:\mathbf{x},\!\mathbf{y}\!\in\!\!\{0,1\}^{2^{2k}} 
 \textrm{  and   } DISJ_{2^{2k}}(\mathbf{x},\mathbf{y})=1
\:\:\Bigl\}
\end{displaymath}
where $(\mathbf{x}\#\mathbf{y}\#\mathbf{x}\#)^{2^k}$ means $2^k$ 
times the concatenation of $\mathbf{x}\#\mathbf{y}\#\mathbf{x}\#$.
\end{definition}
The idea behind the above definition is that, as $\sqrt{2^{2k}}=2^k$ rounds are needed in the worst case for the quantum
protocol computing $DISJ_{2^{2k}}$ presented in \cite{Buhrman+STOC98}, we concatenate the inputs $2^k$ times.

\subsection{Quantum upper bound}

We now show that the language $L_{DISJ}$ can be recognized 
using logarithm space by a bounded-error quantum online machine. 
Actually, we first give the following stronger theorem that states that
the complement of the language $L_{DISJ}$, denoted $\overline{L_{DISJ}}$ and
defined as  $\overline{L_{DISJ}}=\{0,1,\#\}^{\ast}\backslash L_{DISJ}$, 
can be recognized using logarithm space by a one-sided-error quantum online machine.

\begin{theorem}\label{theorem:quantum}
$\overline{L_{DISJ}}\in OQRL$. 
\end{theorem}
This theorem implies our main upper bound.
\begin{corollary}
$L_{DISJ}\in OQBPL$.
\end{corollary} 
\begin{proof2}
We describe a quantum online algorithm using logarithmic space recognizing $L_{DISJ}$,
such that on an input string $w\in\{0,1,\#\}^\ast$:
(1) if $w\in L_{DISJ}$ then the algorithm accepts $w$ with probability 1;
(2) if $w\not\in L_{DISJ}$ then the algorithm rejects $w$ with probability at least $1/4$. 
It is straightforward to convert this algorithm into an OPTM accepting 
the language $\overline{L_{DISJ}}$ and satisfying the conditions of 
Definition \ref{defbounded}.

The most difficult case is when the input satisfies the following three conditions
(if the input does not satisfy one of them, it should be rejected).
\begin{enumerate}
\item[(i)]
The input string is of the form 
\begin{displaymath}
1^k\#\mathbf{x^{(1)}}\# \mathbf{y^{(1)}}\# \mathbf{z^{(1)}}\#\cdots\# \mathbf{x^{(2^k)}}\# 
\mathbf{y^{(2^k)}}\#\mathbf{z^{(2^k)}}\#
\end{displaymath}
where $k\ge 1$, $\mathbf{x^{(i)}},\mathbf{y^{(i)}},\mathbf{z^{(i)}}\in\{0,1\}^{2^{2k}}$ 
are strings in $\{0,1\}^{2^{2k}}$ for all $i\in\{1,\ldots,2^k\}$; and 
\item[(ii)]
$
\mathbf{x^{(1)}=\mathbf{z^{(1)}}=\mathbf{x^{(2)}}}=\mathbf{z^{(2)}}=\cdots=
\mathbf{x^{(2^k)}}=\mathbf{z^{(2^k)}} 
$; and
\item[(iii)]
$
\mathbf{y^{(1)}}=\mathbf{y^{(2)}}=\cdots=\mathbf{y^{(2^k+1)}}
$.
\end{enumerate}

We will present three procedures used to construct our quantum algorithm:
\begin{itemize}
\item
a deterministic classical online procedure $A_1$ that outputs, using logarithm space, $1$ if condition (i)  holds and 
outputs 0 if condition (i) does not hold.
\item
a one-sided-error classical online procedure $A_2$ that outputs,  using logarithm space, when the input satisfies condition (i): 
\begin{itemize}
\item
1 with probability 1 if the input satisfies both (ii) and (iii); and 
\item
0 with probability at least $1/4$ if the input does not satisfies (ii) or (iii).
\end{itemize}
\item
a one-sided-error quantum online procedure $A_3$ that outputs, using logarithm (classical and quantum) space, 
when the input satisfies the three conditions (i), (ii) and (iii): 
\begin{itemize}
\item
$1$ with probability 1 if the input is in $L_{DISJ}$; and 
\item
$0$ with probability at least $1/4$ if the input is not in $L_{DISJ}$.
\end{itemize}
\end{itemize}

The global quantum algorithm runs in parallel the three procedures $A_1$, $A_2$ and $A_3$ and recognizes the input according to the following rule:
if $A_1$ outputs 0, reject; if $A_1$ outputs 1 and $A_2$ outputs 0, reject; if both $A_1$ and $A_2$ output 1, then
accept if $A_3$ outputs 1, reject if $A_3$ outputs 0.
The global algorithm itself has the claimed success probability from the assumptions on $A_1$, $A_2$ and $A_3$: 
if the input is in $L_{DISJ}$, the input accepted with probability 1; else the input is rejected with probability at least
$1/4$.
This algorithm uses logarithm classical and quantum space.

The procedure $A_1$ can be easily implemented.
The rest of this proof consists in describing the two procedures $A_2$ and $A_3$.\vspace{2mm}


\noindent{\bf Procedure $\mathbf{A_2}$}\\
The procedure $A_2$ uses a well-known one-sided-error communication 
complexity protocol for string non-equality \cite{Kush+97}. More precisely,
for any $\mathbf{w}=(w_0,\ldots,w_{2^{2k}-1})\in\{0,1\}^{2^{2k}}$, let us consider the following polynomial in $X$. 
$$F_\mathbf{w}(X)=\sum_{i=0}^{2^{2k}-1}w_iX^i \:\:\textrm{mod } p,$$
where $p$ is an arbitrary prime such that $2^{4k} <p< 2^{4k+1}$. Notice that such a prime is guaranteed to 
exist. Although more efficient techniques exist, the naive strategy consisting in trying all the numbers between 
$2^{4k}$ and $2^{4k+1}$ is sufficient in our case. 
The procedure $A_2$ takes a random integer $t$ in $\Set{0}{p-1}$ and does the following while reading its
input.
\begin{itemize}
\item
For $i=1$ to $2^k$, compute $F_{\mathbf{x^{(i)}}}(t)$, $F_{\mathbf{z^{(i)}}}(t)$
and check whether $F_{\mathbf{x^{(i)}}}(t)=F_{\mathbf{z^{(i)}}}(t)$.
\item
For $i=1$ to $2^k-1$, compute $F_{\mathbf{x^{(i)}}}(t)$, $F_{\mathbf{y^{(i)}}}(t)$, $F_{\mathbf{x^{(i+1)}}}(t)$ and $F_{\mathbf{y^{(i+1)}}}(t)$ and check whether
 $F_{\mathbf{x^{(i)}}}(t)=F_{\mathbf{x^{(i+1)}}}(t)$ and $F_{\mathbf{y^{(i)}}}(t)=F_{\mathbf{y^{(i+1)}}}(t)$.
\end{itemize}
The procedure outputs 1 if all the above tests succeed and outputs 0 if at least one of the tests fails, and
uses $O(k)$ memory bits
On an input satisfying condition (i),
if both (ii) and (iii) holds, then all the tests succeed with probability 1; 
if (ii) or (iii) does not hold, then it can be shown (see for example \cite{Kush+97}) that 
there is at least one test that cannot succeed with probability (on the choice of $t$) greater than $1/2^{2k}$,
and
thus the procedure outputs 0 with probability at least $1-1/2^{2k}$.  \vspace{2mm}  

\noindent{\bf Procedure $\mathbf{A_3}$}\\
We now present the procedure $A_3$ that decides, when the three conditions (i), (ii) and (iii) hold, whether the input string is in $L_{DISJ}$ or not.
Notice that, in this case, the string is in $L_{DISJ}$ if and only if 
$DISJ_{2^{2k}}(\mathbf{x^{(1)}},\mathbf{y^{(1)}})=1$.
The idea of procedure $A_3$ is simply to simulate the communication protocol of Buhrman, Cleve and Wigderson \cite{Buhrman+STOC98}.
We now carefully explain this simulation. 

We first define states and operators that are used by the procedure. 
Let $\ket{\varphi_{k}}$ be the state
$$
\ket{\varphi_{k}}=\frac{1}{2^k}\sum_{i=0}^{2^{2k}-1}\ket{i}
\ket{0}\ket{0}.
$$
Next, define a unitary transformation $S_k$ and, for any string $\mathbf{x}\in\{0,1\}^{2^{2k}}$,
two unitary transformations $V_{\mathbf{x}}$ and $W_{\mathbf{x}}$ as follows.
\begin{eqnarray*}
S_k: \ket{i}\ket{h}\ket{l} &\longmapsto& 
\left\{\begin{array}{cc}
-\ket{i}\ket{h}\ket{l}& \textrm{ if } i\neq 0\\
\ket{i}\ket{h}\ket{l}& \textrm{ if } i= 0
\end{array}\right.\\ 
V_{\mathbf{x}}: \ket{i}\ket{h}\ket{l} &\longmapsto& \ket{i}\ket{h\oplus x_i}\ket{l} \\
W_{\mathbf{x}}: \ket{i}\ket{h}\ket{l} &\longmapsto& (-1)^{h\land x_i}\ket{i}\ket{h}\ket{l}
\end{eqnarray*}
for any $i\in\Set{0}{2^{2k}-1}$ and $h,l\in\{0,1\}$.
Furthermore, define $U_{k}= H^{\otimes 2k}\otimes I_1 \otimes I_1,$
where $H$ is the Hadamard gate and $I_1$ the one-qubit identity operator.
Finally, for any string $\mathbf{x}\in\{0,1\}^{2^{2k}}$, we denote by $R_\mathbf{x}$ the following unitary transformation.
\begin{displaymath}
R_\mathbf{x}:\ket{i}\ket{h}\ket{l}\longmapsto\ket{i}\ket{h}\ket{l\oplus(h\land x_{i})},
\end{displaymath}
for all $i\in\Set{0}{2^{2k}-1}$ and $h,l\in\{0,1\}$.

We are now ready to present procedure $A_3$.

\begin{enumerate}
\item
$\ket{\varphi}\gets \ket{\varphi_{k}}$.
\item
Take an integer $j$ chosen at random according to the uniform distribution over $\Set{0}{2^k-1}$.
\item
For $i$ from 1 to $j$, do
$\ket{\varphi}\gets U_{k}S_{k}U_{k}V_{\mathbf{z^{(i)}}} W_{\mathbf{y^{(i)}}}V_{\mathbf{x^{(i)}}}\ket{\varphi}.$
\item
$\ket{\varphi}\gets R_{\mathbf{y^{(j+1)}}}V_{\mathbf{x^{(j+1)}}}\ket{\varphi}$.
\item
Measure the last qubit of $\ket{\varphi}$. Let $b$ be the outcome. Then output $(1-b)$.
\end{enumerate}
The application of the gates $U_k$, $S_k$, $V_{\mathbf{x^{(i)}}}$, 
$W_{\mathbf{y^{(i)}}}$, $V_{\mathbf{z^{(i)}}}$ and $R_{\mathbf{y^{(i)}}}$ can be 
done easily while
reading the inputs $\mathbf{x^{(i)}}$, $\mathbf{y^{(i)}}$ and $\mathbf{z^{(i)}}$
(i.e., a quantum circuit implementing them can be computed using only $O(k)$ classical work space, as required by 
Definition \ref{defbounded}). 
From the equality 
$$V_{\mathbf{x}} W_{\mathbf{y}}V_{\mathbf{x}}\left( 
\sum_{i=0}^{2^{2k}-1}\alpha_i\ket{i}\ket{0}\ket{0}\right)=
\sum_{i=0}^{2^{2k}-1}\alpha_i(-1)^{x_i\land y_i}\ket{i}\ket{0}\ket{0},$$ 
we see that each application of the loop 3 corresponds to one iteration of Grover's algorithm.
The state at the end of step 4 will be  
$$\sum_{i=0}^{2^{2k}-1}\beta_i\ket{i}\ket{0}\ket{x_i\land y_i}$$ 
for some amplitudes $\beta_i$, and corresponds to to the state obtained after $j$ iterations of Grover's algorithm.
If the value $DISJ_{2^{2k}}(\mathbf{x^{(1)}},\mathbf{y^{(1)}})$ is $1$, then measuring the last register gives
$0$ with probability 1. Then the procedure outputs $1$ with probability 1. 
Otherwise denote by $t$ the number of coordinates such that $x_i\land y_i=1$.
If $t=2^{2k}$, then the procedure always outputs $1$ so let us suppose that $0<t<2^{2k}$.
Denote by $\theta$ the angle such that $\sin^2\theta=t/2^{2k}$, with $0<\theta<\pi/2$. 
The analysis by Boyer, Brassard, H{\o}yer and Tapp \cite{Boyer+98} shows that the probability of measuring 1 
(and thus outputting 0) is
$$
\frac{1}{2}-\frac{\sin(4\cdot 2^k\theta)}{4\cdot 2^k\sin(2\theta)}\ge \frac{1}{4}.
$$
Thus the procedure $A_3$ outputs, when the input satisfies condition (i), (ii) and (iii), 1 with probability 
$1$ if the input is in $L_{DISJ}$ and outputs 0 with probability at least $1/4$ if the input is not in $L_{DISJ}$.
\end{proof2}
\subsection{Classical lower bound}

We now show that any classical OPTM recognizing
$L_{DISJ}$ with bounded-error has to use $\Omega(n^{1/3})$ work space, where $n$ is the input length.

\begin{theorem}
There exists a constant $c_1>  0$ such that $$L_{DISJ}\notin OBPSPACE(c_1n^{1/3}).$$ 
\end{theorem}
\begin{proof}
Our proof is inspired by an argument by Ablayev \cite{AblayevTCS96}.
Consider an OPTM $M$ recognizing the language $L_{DISJ}$ with bounded-error.
Suppose, without loss of generality, that $M$ does not halt before
having read all the input.

We show how to convert the computation of $M$ on inputs of the form 
\begin{equation}\label{eq:DISJlow}
1^k\#
(\mathbf{x}\#\mathbf{y}\#\mathbf{x}\#)^{2^k} 
\:\textrm{ with }\:
\mathbf{x},\mathbf{y}\in\{0,1\}^{2^{2k}}
\end{equation}
into a communication protocol 
that computes $DISJ_{2^{2k}}$. Let $\mathbf{x}$ be Alice's input and $\mathbf{y}$ Bob's input.
Given two configurations $C_1$ and $C_2$ of $M$,
we write $C_1\xrightarrow{\mathbf{w}} C_2$ if the configuration $C_2$ is reachable from $C_1$ on 
the word $\mathbf{w}$ with positive probability, with the additional condition that the last computational step before reaching $C_2$ includes
a move of the head of the input tape to the right of the last character of $\mathbf{w}$ (i.e., $\mathbf{w}$ has been 
completely read and the next character on its
right is for the first time scanned). 
Denote by $C^{(0)}$ the initial configuration of $M$.

Alice and Bob use the following protocol to compute the value
$DISJ_{2^{2k}}(\mathbf{x},\mathbf{y})$. 
\begin{enumerate}
\item
For $i$ from 1 to $3\cdot 2^k-1$, do the following (if the protocol has not been interrupted):\vspace{1mm}

{\bf Step number $\mathbf{i}$:} Alice's turn if $i \not\equiv 2\bmod 3$, Bob's turn else.\\ 
The player computes all $C_j$ such that $C^{(i-1)}\xrightarrow{\mathbf{w}}C_j$ and their probabilities $p_j$,
where 
$$
\mathbf{w}=\left\{\begin{array}{cl}
1^k\#\mathbf{x}\#&\textrm{ if } i=1\\ 
\mathbf{x}\#&\textrm{ if } i\neq 1 \textrm{ and }i \not\equiv 2\bmod 3\\
\mathbf{y}\#&\textrm{ if } i\neq 1 \textrm{ and }i \equiv 2\bmod 3
\end{array}
\right.
$$
The player sends a message consisting of the configuration $C_j$ with probability $p_j$. In this case, denote by $C^{(i)}$ the configuration sent.
With probability $1-\sum_{j}p_j$, the player stops the protocol and outputs $0$ (this corresponds to the case of infinite work of the machine). 
Notice that  if $i \equiv 0\bmod 3$ (Alice's turn), Alice being the next player, she does not really need to send the configuration.
For the simplicity of the analysis, we will suppose that even in this case Alice sends a message (to herself).
\item
Alice computes the probability $p$ that the machine halts on an accepting configuration from the configuration 
$C^{(3\cdot 2^k-1)}$ on the input $\mathbf{x}\#$.
She outputs $1$ with probability $p$ and outputs $0$ with probability $1-p$.
\end{enumerate}

The condition that $M$ recognizes $L_{DISJ}$ with bounded error implies that the above protocol 
is a correct bounded-error communication protocol for $DISJ_{2^{2k}}$. 
Let $\mathscr{C}^{(i)}_{k}$ be the set of configurations that are sent with positive probability
at step $i$ for at least one input of the form (\ref{eq:DISJlow}).
The number of bits exchanged by the protocol is less than $$\sum_{i=1}^{3\cdot 2^k-1}\ceil{\log\abs{\mathscr{C}_{k}^{(i)}}}.$$
This implies that there exists necessarily an integer $i_0\in\Set{1}{3\cdot 2^k\!-\!1}$ such that 
$$\ceil{\log\abs{\mathscr{C}_{k}^{(i_0)}}}\ge \frac{1}{3\cdot2^k-1}\sum_{i=1}^{3\cdot 2^k-1}\ceil{\log\abs{\mathscr{C}_{k}^{(i)}}}.$$
From Theorem \ref{theorem:lowerbound}, the number of bits exchanged by the protocol is
necessary $\Omega(2^{2k})$. Thus $\log\abs{\mathscr{C}^{(i_0)}_{k}}$ is at least
$\Omega(2^{2k}/(3\cdot 2^k-1))=\Omega(2^k)$.
From Fact \ref{fact3}, the space necessarily for $M$ is thus $\Omega(2^k)=\Omega(n^{1/3})$, where 
$n=\Theta(2^{3k})$ is the size of the inputs of the form (\ref{eq:DISJlow}). 
\end{proof}

This lower bound is essentially tight, as shown in the next proposition.
\begin{proposition}
There exists a constant $c_2$ such that $L_{DISJ}\in OBPSPACE(c_2n^{1/3})$.
\end{proposition}
\begin{proof}
Suppose that the input is $1^k\#(\mathbf{x}\#\mathbf{y}\#\mathbf{x}\#)^{2^k}$ with $\mathbf{x},\mathbf{y}\in\{0,1\}^{2^{2k}}$.
The other cases can be dealt with by using the same classical techniques as in the algorithm of Theorem \ref{theorem:quantum}.
Now, we
decompose  $\mathbf{x}$ into $2^k$ blocks of $2^k$ bits denoted $\mathbf{[x]_1},\ldots,\mathbf{[x]_{2^k}}$, and decompose 
$\mathbf{y}$ similarly.
The OPTM checks successively the value of 
$DISJ_{2^k}(\mathbf{[x]_i},\mathbf{[y]_i)}$, for $i$ from $1$ to $2^k$,
using the trivial procedure consisting of keeping all the bits of the block $\mathbf{[x]_i}$ in memory before matching them with the bits of $\mathbf{[y]_i}$.
This uses $O(2^k)=O(n^{1/3})$ space, where $n$ is the input size.  
\end{proof}
\section*{Acknowledgments}
The author is grateful to Hiroshi Imai, Seiichiro Tani and Tomoyuki Yamakami for helpful comments about this work.



\end{document}